\documentclass[12pt]{article}\usepackage{amsmath,amsfonts}
\makeatletter \@addtoreset{equation}{section}

\makeatother
\begin{document}
\baselineskip 18pt%
\begin{titlepage}
\vspace*{1mm}%
\hfill%
\vbox{
    \halign{#\hfil        \cr
           IPM/P-2007/050 \cr
           SUT-P-07-2b   \cr
                     } 
      }  
\vspace*{15mm}%

\centerline{{\Large {\bf  Double-Horizon Limit, AdS Geometry
  }}} \centerline{{\Large {\bf and Entropy Function  }}}

\vspace*{5mm}
\begin{center}
{ H. Arfaei, R. Fareghbal}%

\vspace*{0.8cm}

{\it {Department of Physics, Sharif University of Technology\\
P.O.Box 11365-9161, Tehran, IRAN}}\\
{\it {And}}\\
{\it {Institute for Studies in Theoretical Physics and Mathematics (IPM)\\
P.O.Box 19395-5531, Tehran, IRAN}}\\

{E-mails: {\tt arfaei@mail.ipm.ir, fareghbal@theory.ipm.ac.ir}}%
\vspace*{1.5cm}
\end{center}

\begin{center}{\bf Abstract}\end{center}
\begin{quote}
We start  from a generic metric which describes   four dimensional
stationary black holes in an arbitrary theory of gravity and show
that the $AdS_2$ part of the near horizon geometry is a consequence
of the double-horizon limit and finiteness .  We also show that
 the field configurations  of the near horizon are determined if the same
conditions are applied to the equations of motion. This is done by
 showing that in the double-horizon limit  field equations at the horizon decouple from the
bulk of the space. Solving these equations gives the near horizon
field configurations. It is shown that these decoupled equations can
be obtained from an action derived from the original action by
applying the double-horizon condition. Our results agree with the
entropy function method.

\end{quote}
\end{titlepage}

\section{Introduction }
It is well known  that in a large class of the gravity theories
coupled to  a number of  scalars and gauge fields, the value of the
scalars at the horizon of the extremal black holes is independent of
the values at the large distance. This phenomenon called attractor
mechanism, has been first shown  for the supersymmetric theories
\cite{Ferrara:1995ih}-\cite{Ferrara:1996dd} and was later  studied
and proved  for the non-supersymmetric cases
\cite{Goldstein:2005hq}.
 Application of
the attractor mechanism  in string theory and its importance in the
counting of the states of black holes is disscussed in
\cite{Astefanesei:2006sy} and \cite{Dabholkar:2006tb}.

A remarkable progress in  understanding of the  attractor mechanism
came with the works of  {\it Sen} who introduced the {\it entropy
function} method \cite{Sen:2005wa}-\cite{Sen:2005iz}. He showed that
not only scalar fields but all parameters of the near horizon of an
extremal black hole is fixed by extremizing a function, called the
entropy function, which is evaluated near the horizon.   Sen's
method resulted in a generalized attractor mechanism and was shown
to work in a number of cases.

 {In the entropy function method,
 one starts with characterization   of the extremal black holes with their
 near horizon geometry. It is taken to have  an  $AdS_2$ factor. We would like to
 address the question of deeper  physical properties that leads to
 such geometric characterization in most known cases. It is also of
 our interest to examine whether one can find similar method for the non-extremal
 cases which do not have $AdS_2$ space  in
 their  near horizon geometry }. There has been several attempts  to
answer these questions \cite{Arfaei:2006qr}-
\cite{Astefanesei:2007bf}.

In this paper we show that both
   questions can be
approached  from a single  point of view which is imposing
finiteness  on the physical quantities   at the horizon of the
double-horizon black holes.  For most  of the known four dimensional
black hole solutions, extremality coincides with the double-horizon
limit, i.e. the radii of the event inner and outer horizons
coincide. This results in zero surface gravity on the horizon and
zero temperature for the black hole. Moreover, the singularities at
the horizons are coordinate singularities and can be removed  by
coordinate transformation, thus we expect that the scalars
constructed by the metric are finite at the horizons.

The physical reason for the attractor mechanism is the infinite
distance to the double horizon that prevents the information to
affect the horizon physics \cite{Kallosh:2006bt}.
 This distance from an arbitrary point to a simple  horizon is finite
and hence allows the bulk information  reach the horizon and
therefore blocks the attractor mechanism. The double-horizon
property is essential in the divergence of the distance in the first
and finiteness of the latter.

This work follows and completes the work done in
\cite{Arfaei:2006qr}. In section two, we start from a generic metric
which describes four dimensional stationary black holes. This form
is completely general and does not depend on any particular theory.
Then we explore the consequences of the finiteness of scalars
constructed from the metric. The finiteness is justified since the
singularities of the horizon must be removable. We demand the
finiteness of $R$, $R_{\mu\nu}R^{\mu\nu}$ and
$R_{\mu\nu\rho\sigma}R^{\mu\nu\rho\sigma}$  and find certain
restrictions on the components of the metric. The double-horizon
condition imposes further restrictions. Using these restrictions and
following the method of \cite{Bardeen:1999px}, we obtain  the near
horizon geometry of the stationary black holes which is the starting
point of \cite{Astefanesei:2006dd} in generalizing the entropy
function method to the rotating black holes. Our result clearly
shows that the $AdS$ near horizon geometry is a consequence of the
finiteness at the horizon and the double-horizon limit.

In  section  three we use  our results in \cite{Arfaei:2006qr} and
the  finiteness assumption and the double-horizon condition on the
equations of motion to cast  the equations to a   set of equations
that are decoupled from the bulk.  We do this by starting from the
general ansatz for the metric which describes both  the extremal and
non-extremal black holes. The decoupling occurs due to vanishing of
all  r-derivative terms. First we show  the decoupling of the
dynamics in Einstein gravity coupled to a number of scalars and
abelian gauge fields and then extend it to  $f(R)$ gravities.
  For $f(R)$ gravities we use the technique  of the
equivalence of these theories with Einstein-Scalar gravity
\cite{Maeda:1988ab}. These equations are non-linear ordinary
differential equations with respect to $\theta$. The solution(s) to
these equations specifies the field configuration(s) at the near
horizon. However concluding the attractor mechanism from this
decoupling   is possible if we prove that our equations have a
unique solution.

 Our analysis also shows that if we do not apply the
double-horizon limit, the field equations at the horizon do not
decouple from the bulk and thus we can not expect an attractor
mechanism for the case of simple horizon. In particular we observe
that it is not possible to reproduce the dynamics on the simple
horizon  by extremizing a function which is only defined at the
horizon.

Recently  \cite{Cai:2007ik}-\cite{Garousi:2007nn} has found examples
of non-extremal black holes for which the entropy function method
works. These results  are not in contradiction with ours, since the
original metric which describes black holes in those cases are not
the same as one we consider in this work.

Our method for  dealing with the rotating black holes is the
reduction of $\phi$-coordinate in Kaluza-Klein style. Because of
axisymmetry of the stationary black holes, parameters of the metric
do not have $\phi$ dependence. Moreover $\phi$ is a periodic
coordinate, therefore similar to  the Kaluza-Klein reduction, we can
reduce this direction and find a three dimensional gravity theory in
which apart from the original scalar and gauge fields additional
scalar and gauge fields origination from componenets of four
dimensional metric and fields exist. The interesting point is that
the charge of the new gauge field is exactly the angular momentum of
the four dimensional black hole. This technique simplifies the
derivation and the form of the decoupled equations.

In  section four we establish the relation of the equations to the
entropy function method.  We  reproduce the decoupled equations by
extremizing a function  at the horizon. The difference with entropy
function method is that this function is calculated using the
general form for the fields of the theory and imposing the
double-horizon and finiteness conditions.

 This results show that the possibility of derivation
 of the equations of motion by extremizing a function at
the horizon, is a direct consequence of the double-horizon limit.
 We also  investigate the result of our assumed conditions on the
  the Wald's entropy formula. Following the method of
 \cite{Astefanesei:2006dd} we find
the results previously obtained by the entropy function method. The
authors of  reference \cite{Cai:2007an} have used the zero surface
gravity property of the extremal black holes  to simplify the Wald
formula for static solutions. Their results are in agreement with
our proposal that the decoupled physics of the horizon is a
consequence of the double-horizon limit.

\section{ Double-Horizon limit and AdS geometry}
  We consider $4$-dimensional
stationary black holes with axial symmetry. They are described by
the generic metric
\begin{equation}\label{general form}
    ds^2=-a(r,\theta)\,dt^2+\frac{b(r,\theta)}{S(r)}\,dr^2+e(r,\theta)\,d\theta^2+f(r,\theta)\,d\phi^2+2c(r,\theta)\,dt\,d\phi
\end{equation}
where $a(r,\theta)$, $b(r,\theta)$, $c(r,\theta)$, $e(r,\theta)$ and
$f(r,\theta)$ are assumed to be regular functions. Event horizons
are located at $r=r_H$ where $S(r_H)=0$. In writing this metric, it
is assumed that event horizons are Killing horizons too. This is a
general form based on the assumed symmetries. At this stage we
consider only the symmetry which is a reflection of the topology of
the horizon in four dimension and no particular theory of the
gravity is assumed in this part. Therefore the result is applicable
to a wide class of
 theories . We consider the cases
that $S(r)$ has non-zero roots and exclude the naked singularities.

 Let us first impose the finiteness assumption  on
the determinant of this metric. The determinant $g$ is a
coordinate dependent quantity. The quantity which is invariant and
coordinate independent is $\Delta V=\sqrt{-g}\,d^4x$. If we assume
that the volume of any finite neighborhood near the horizon is
finite then from finiteness of $d^4x$ we conclude that $\sqrt{-g}$
is also finite. The determinant is,
\begin{equation}\label{determinant}
    g=-\frac{b(r,\theta)\,e(r,\theta)\Big(a(r,\theta)f(r,\theta)+c(r,\theta)^2\Big)}{S(r)}
\end{equation}
  Finiteness  requires,
\begin{equation}\label{finiteness of g}
    a(r,\theta)\,f(r,\theta)+c(r,\theta)^2=\,S(r)\,v(r,\theta)
\end{equation}
where $v(r,\theta)$ is a regular  function.

 Finding $a(r,\theta)$ from (\ref{finiteness of g}) and
substituting it in (\ref{general form}) and redefining new regular
functions, we obtain,
\begin{equation}\label{metric simplifyed 2}
    ds^2=A(r,\theta)\Bigg(-\frac{S(r)}{B(r,\theta)}\,dt^2+\frac{dr^2}{S(r)}\Bigg)+E(r,\theta)\,d\theta^2+F(r,\theta)\Bigg(d\phi+C(r,\theta)\,dt\Bigg)^2
\end{equation}

Using this metric we can calculate scalars such as $R$,
$R_{\mu\nu}R^{\mu\nu}$ and
$R_{\mu\nu\rho\sigma}R^{\mu\nu\rho\sigma}$. Assuming that the
singularities of the horizons are coordinate singularities, we
conclude that  these scalars are all  finite there. However,
computations of these scalars  show that they include  terms that
have $S(r)$ factor in their denominator which will diverge unless
the nominators also develop similar factors.
   Investigation of all these terms
   shows that some  have $\frac{\partial}{\partial\theta}B(r,\theta)$ and some  $\frac{\partial}{\partial\theta}C(r,\theta)$ factor in their nominator.
   If these derivatives are  proportional to $S(r)$ then all of them  become finite at the horizons. Thus  finiteness imposes   the  conditions,
\begin{equation}\label{1finitness of C anb B}
        \frac{\partial}{\partial\theta}C(r,\theta)\Bigg\vert_{r=r_H}\!\!\!\!\!\!\!\!\!\!\!\!=0,\qquad\frac{\partial}{\partial\theta}B(r,\theta)\Bigg\vert_{r=r_H}\!\!\!\!\!\!\!\!\!\!\!\!=0
 \end{equation}
This means that these two functions $B$ and $C$ must have the
forms,
\begin{equation}\label{form of B}
    B(r,\theta)=S(r)B_1(r,\theta)+B_2(r)
\end{equation}
\begin{equation}\label{form of C}
    C(r,\theta)=S(r)C_1(r,\theta)+C_2(r)
\end{equation}
For our argument it is sufficient to take  the first power of $S(r)$
in the right hand side. One can absorb higher powers in the
functions $B_1(r,\theta)$ and $C_1(r,\theta)$. The  regularity of
the derivatives of  $B(r,\theta)$ and $C(r,\theta)$ at the horizon
do not allow that power of $S(r)$ be less than one.
 It is obvious that at the horizons $B(r,\theta)$ and $C(r,\theta)$ are
constants and do not have $\theta$-dependence. We show their values
 at the outer horizon by $B$ and $C$.

For black holes with the double-horizon  where the inner and outer
horizons coincide , we have
\begin{equation}
    S(r)=(r-r_H)^2
\end{equation}
where $r_H$ is the radius of the horizons.

Using (\ref{form of C}) and applying the double-horizon condition,
it is not difficult to see that $\alpha$  defined as
\begin{equation}\label{definition of alpha}
    \alpha\equiv\frac{\partial}{\partial r}C(r,\theta)\Bigg\vert_{r=r_H}\!\!\!\!\!\!\!\!\!
 \end{equation}
 is a constant .

Now we are ready  to find the near horizon geometry of the
double-horizon black holes. First, we expand the functions
specifying  the metric around  $r_H$ in terms of $ r-r_H $. Using
(\ref{1finitness of C anb B}) and (\ref{definition of alpha}), we
get
\begin{eqnarray}\label{expanding}
                                                          \nonumber A(r,\theta) &\approx& A(\theta)+\frac{\partial A}{\partial r}\Bigg\vert_{r=r_H}\!\!\!\!\!\!\!\!\!(r-r_H)
                                                          ,\\
                                                         \nonumber B(r,\theta) &\approx& B+\frac{\partial B}{\partial r}\Bigg\vert_{r=r_H}\!\!\!\!\!\!\!\!\!(r-r_H)
                                                         ,\\
                                                        \nonumber  C(r,\theta) &\approx& C+\alpha(r-r_H)
                                                        ,\\
                                                        \nonumber  E(r,\theta) &\approx& E(\theta)+\frac{\partial E}{\partial r}\Bigg\vert_{r=r_H}\!\!\!\!\!\!\!\!\!(r-r_H)
                                                        ,\\
                                                          F(r,\theta) &\approx& F(\theta)+\frac{\partial F}{\partial
                                                          r}\Bigg\vert_{r=r_H}\!\!\!\!\!\!\!\!\!(r-r_H).
                                                        \end{eqnarray}
The near horizon coordinates $(\hat t,\hat r, \theta,\hat \phi)$ are
defined as \cite{Bardeen:1999px};
\begin{eqnarray}\label{near horizon coordinate}
 \nonumber t &=& \frac{\sqrt B}{\lambda}\,\hat t \\
\nonumber  r &=& r_H+\lambda \hat r \\
  \phi &=& \hat \phi-C\frac{\sqrt B}{\lambda}\hat t
\end{eqnarray}
in the limit $\lambda\to 0 $.

Substituting (\ref{expanding}) in (\ref{metric simplifyed 2}) and
using (\ref{near horizon coordinate}), we obtain the metric in terms
of the near horizon coordinates,
\begin{equation}\label{near horizon geometry of 4d}
    ds^2=A(\theta)\Bigg(-\hat r^2\,d\hat t^2+\frac{d\hat r^2}{\hat r^2}\Bigg)+E(\theta)\,d\theta^2+F(\theta)\Bigg(d\hat\phi+\Sigma \hat r d\hat t\Bigg)^2
\end{equation}
where
\begin{equation}
    \Sigma=\sqrt B \alpha\
\end{equation}
In this coordinate system which is suitable for  the near horizon,
$g_{\theta\theta}$ component of the  metric has only $\theta$
dependence. This results a reparametrization freedom which allows us
to fix $g_{\theta\theta}$ up to a constant by defining $\hat\theta$
as
\begin{equation}\label{def of hattheta}
    \hat\theta=\frac{1}{\Gamma}\,\int_0^\theta\,d\theta^\prime\sqrt{E(\theta^\prime)}
\end{equation}
where
\begin{equation}\label{def of gamma}
    \Gamma=\frac{1}{\pi}\int_0^\pi\,d\theta^\prime\sqrt{E(\theta^\prime)}
\end{equation}
It is clear that $\hat\theta$ goes between $0$ to $\pi$. Near
horizon geometry finally takes the form
\begin{equation}\label{near horizon geometry of 4d-final form}
    ds^2=A(\hat\theta)\Bigg(-\hat r^2\,d\hat t^2+\frac{d\hat r^2}{\hat r^2}\Bigg)+\Gamma^2 d\hat\theta^2+F(\hat\theta)\Bigg(d\hat\phi+\Sigma \hat r d\hat t\Bigg)^2
\end{equation}

  The main  point  is that (\ref{near horizon geometry of 4d-final form}) is a result of the
finiteness  at the horizon and the double-horizon condition. This
result is general and  does not depend on the action of the theory.
It is also independent of the asymptotic behavior of the metric.
Hence it can be applied even to metrics embedded  in non
asymptotically flat space. (\ref{near horizon geometry of 4d-final
form}) is the starting point of \cite{Astefanesei:2006dd} in
generalizing entropy function method for the rotating black holes.
In \cite{Astefanesei:2006dd} it is taken as the beginning point for
the rest of the  argument but we have shown it as a result of
physically simple assumptions. This shows that $AdS_2$ near horizon
geometry which is used in entropy function method as the definition
of the extremal black holes is a consequence of the finiteness and
the  double-horizon.  In the next sections we will show that in
order to find the parameters of the near horizon it is not necessary
to go to the near-horizon geometry and apply entropy function
method. Starting from the metric (\ref{metric simplifyed 2}) and
writing the equations at the horizon and applying the double-horizon
condition, we can find the parameters of the near horizon geometry
directly.

Another point  is that values of $B$ and $C$ at the horizon are
absorbed in the definition of $\hat t$ and $\hat\phi$. This means
that there is a  reparametrization freedom at the horizon and the
values of $B$ and $C$ at the horizon can not be determined  by
equations of motion and are not physical. We can only find $\Sigma$
which is sufficient for specifying the near horizon geometry. This
will be clear in the next sections where by using double-horizon
condition we get a set of decoupled equations at the horizon for
$A(\hat\theta)$, $F(\hat\theta)$, $\Gamma$ and $\Sigma$.

\section{ At the Horizon of   Double-Horizon Black Holes }
Analysis of the previous section shows the importance of the
finiteness and the double-horizon condition  in  finding the near
horizon geometry of the double-horizon black holes. Furthermore,  in
this section we would like  to show that these assumptions are
enough to fix the field configuration of the  near horizon geometry
. The interesting point here is that this fixing is possible without
directly using the near horizon field configurations. It is done by
applying double-horizon limit on the field equations come from
variation of a generic action and using generic ansatz for  the
field content of the theory . The result applies to a large class of
theories and is independent of the details of the dynamics in the
bulk.

 We first consider rotating  charged black holes in a
theory of gravity with scalars $\Phi_I\quad \!\!(I=1,2,...)$ and
abelian gauge fields $A^{(K)}\quad \!\!(K=1,2,...) $ described by
the
 action\begin{equation}\label{action }
   S=\frac{1}{\kappa^2}\int d^4 x
    \sqrt{-\hat g}\Big(\hat R-h_{IJ}(\Phi)\partial_\mu\Phi_I\partial^\mu\Phi_J-w_{KL}(\Phi) F^{(K)}_{\mu\nu} F^{(L)\,\mu\nu}-V(\Phi)\Big)
\end{equation}
where $w_{KL}(\Phi)$ determines the coupling of the scalars to the
gauge fields and $V(\Phi)$ is a potential term for the scalars.
After analyzing this case we will consider more general theories.
 As shown in the previous section the form of the metric is
\begin{equation}\label{ansatz for the metric charged}
  d\hat
  s^2=A(r,\theta)\Bigg(-\frac{S(r)}{B(r,\theta)}dt^2+\frac{1}{S(r)}dr^2\Bigg)+E(r,\theta)\,d\theta^2+
    F(r,\theta)\Bigg(d\phi+C(r,\theta)dt\Bigg)^2
  \end{equation}
 where\begin{equation}
    S(r)=(r-r_+)(r-r_-)
  \end{equation}
$r_+$ and $r_-$ are the radii of the outer and inner horizons. It is
clear that for the case of double-horizon black holes i.e.
$r_+=r_-$, not only $S(r)$ but also $\frac{dS(r)}{dr}$ vanishes at
the horizons and for all cases $\frac{d^2S(r)}{dr^2}=2$.

Using axial symmetry of the rotating stationary black holes, we
choose following ansatz for the  scalars and gauge fields:
\begin{equation}\label{ansatz for the scalar scalar}
    \Phi_I=\Phi_I(r,\theta)
\end{equation}\begin{equation}\label{ansatz for the gauge filed 4d}
    A^{(K)}_\mu\,dx^\mu
   = A^{(K)}_t(r,\theta)dt+ A^{(K)}_\phi(r,\theta)d\phi\end{equation}

It is assumed that this black hole has angular
momentum $J$, electric charges $Q^{(K)}$ and magnetic charges
$P^{(K)}$  defined through the relation
\begin{equation}\label{def of maghnetic charge}
    P^{(K)}=\int\,d\theta d\phi
    F^{(K)}_{\theta\phi}=2\pi\Big(A^{(K)}_\phi(\pi)-A^{(K)}_\phi(0)\Big)
\end{equation}

As we saw in the  previous section,  requirement of the finiteness
of the scalars constructed by the metric imposes conditions
(\ref{1finitness of C anb B}) on the functions $B(r,\theta)$ and
$C(r,\theta)$.

Now we  demand  that at the horizons, not only the scalars
constructed from  the metric but also all the terms that appear in
the action to be finite . It is possible to consider (\ref{action
}) as the KK-reduction of a higher dimensional pure gravity which
only has $\sqrt {-g^{(d)}}\,R^{(d)}$ term. The singularity of the
horizons in this higher dimensional  picture is a coordinate
singularity too and therefor $R^{(d)}$ must be finite. After the
reduction, this term breaks to three terms of the 4d action
(\ref{action }) and thus sum of these terms must be finite at the
horizons. However, all of these terms must be separately finite.
The $R$ term is finite since the singularity is coordinate
singularity. The other two terms are both positive and hence each
of them must be separately finite.
 $h_{IJ}$ and $w_{KL}$ are also finite since are reduced from the higher dimensional regular metric components. Finiteness of $ F_{\mu\nu} F^{\,\mu\nu}$ requires that
\begin{equation}\label{finiteness of ff}
     F^{(K)}_{\theta t}(r,\theta)-C(r,\theta) F^{(K)}_{\theta\phi}(r,\theta)\Bigg\vert_{r=r_H}\!\!\!\!\!\!\!\!\!\!\!\!=0
 \end{equation}
Using (\ref{1finitness of C anb B}) and (\ref{ansatz for the gauge
filed 4d})  this condition takes the form:
\begin{equation}\label{aa}
    \frac{\partial}{\partial \theta}\Big(A^{(K)}_t(r,\theta)-C(r,\theta)A^{(K)}_\phi(r,\theta)\Big)\Bigg\vert_{r=r_H}\!\!\!\!\!\!\!\!\!\!\!\!=0
\end{equation}
 Therefore the  left hand side is proportional to $S(r)$ and we can
 write
\begin{equation}\label{finnitness on AA}
    \frac{\partial}{\partial
    \theta}\Big(A^{(K)}_t(r,\theta)-C(r,\theta)A^{(K)}_\phi(r,\theta)\Big)=S(r)f^{(K)}(r,\theta)
\end{equation}
where $f^{(K)}(r,\theta)$ is a regular   function\footnote{The
power of $S(r)$ in the right hand side of (\ref{finnitness on AA})
can be $n\,(n\geq 1) $. We absorb $S(r)^{n-1}$ in
$f^{(K)}(r,\theta)$. This power can not be less than one since
makes $F_{\mu\nu}F^{\mu\nu}$ infinite.}.
 By integrating both sides with respect to $\theta$ we obtain
 \begin{equation}\label{gaue finite2}
    A^{(K)}_t(r,\theta)-C(r,\theta)A^{(K)}_\phi(r,\theta)=S(r)f^{(K)}_1(r,\theta)+f^{(K)}_2(r)
\end{equation}
 It is obvious from this relation that $\beta^{(K)}$'s  defined as

\begin{equation}\label{definition of beta}
 \beta^{(K)}=\frac{\partial}{\partial
 r}\Big(A^{(K)}_t(r,\theta)-C(r,\theta)A^{(K)}_\phi(r,\theta)\Big)\Bigg\vert_{r=r_H}
 \end{equation}
are  constant for the double-horizon black holes. It is not
difficult to see that in the near horizon coordinate (\ref{near
horizon coordinate}), we have
\begin{equation}\label{Fharrhatt}
    F^{(K)}_{\hat r\hat t}=\sqrt B \,\beta^{(K)}
\end{equation}
thus our conditions  guarantee   that this component of the field
strength is constant  in the near horizon  coordinate.

In dealing with the  rotating black holes it is easier to apply a
reduction similar to KK reduction  in the
 $\phi$-direction. The axisymmetry of the rotating black hole
 background guarantees
 that non of the metric components  depends on $\phi$ which allows such
  reduction. Since
 this direction is  periodic with period of $2\pi$, we can
 look at it as a compact direction with unit  radius of
 compactification.   The four dimensional  rotating black hole  will be
 converted to a non-rotating but charged black object
 represented in this new
 three dimensional picture. This black object is the source of the new
 scalar
 fields and a new gauge fields. Angular momentum of the 4d rotating
 black hole  will be charge of the new 3d black object.

 Let us define
\begin{eqnarray}\label{def of MN charged}
\nonumber  M(r,\theta) &\equiv& F(r,\theta)\,A(r,\theta) \\
  N(r,\theta)  &\equiv& F(r,\theta)\,E(r,\theta)
\end{eqnarray}
Reducing the action (\ref{action }) in the $\phi$-direction and
 changing   to Einstein frame, we obtain
\begin{equation}\label{three dimentinal action charged case}
   \begin{split} S=\frac{2\pi}{\kappa^2}\int d^3 x
    \sqrt{- g}\Big( &R-H^{ij}(\Phi)\partial_\mu\Phi_i\partial^\mu\Phi_j-W_{ab}(\Phi) F\,^{(a)}_{\mu\nu}
    F^{(b)\,\,\mu\nu}-U(\Phi)\Big)\cr\end{split}
\end{equation}
where the three dimensional metric takes  the form,
\begin{equation}\label{3d metric charged case}
     ds^2=M(r,\theta)\Bigg(-\frac{S(r)}{B(r,\theta)}\,dt^2+\frac{dr^2}{S(r)}\Bigg)+N(r,\theta)\,d\theta^2
\end{equation}
The scalars are given as

\begin{equation}\label{scalars}
{\bf\Phi}=\left(\begin{array}{c}
  -\frac{1}{2}lnF(r,\theta) \\ \\
   A^{(K)}_\phi \\ \\
  \Phi_I(r,\theta)
\end{array}\right)
\end{equation}
\\
The first two components are from the $\phi \phi$ metric component
and $ \phi $ component of the gauge potential.

The three dimensional gauge potentials  are
\begin{equation}\label{gaugefilelds}
{\bf A_t}=\left(\begin{array}{c}
  C(r,\theta) \\ \\
  A_t^{(K)}(r,\theta)-C(r,\theta)A_\phi^{(K)}(r,\theta)
\end{array}\right)
\end{equation}
\\
and the moduli metric is
\begin{equation}\label{definition of H}
    {\bf H}=\left (\begin{array}{ccc}2&\qquad 0&\qquad0\\ \\0&\qquad\frac{2}{F(r,\theta)}w_{KL}&\qquad0\\ \\0&\qquad0&\qquad h_{IJ} \end{array} \right)
\end{equation}
\newline
The gauge field couplings are given by
\begin{equation}\label{definition of W}
    {\bf W}={F(r,\theta)}\left (\begin{array}{cc}\frac{1}{4}\,F(r,\theta)+w_{KL}A^{(K)}_\phi A^{(L)}_\phi&\qquad\qquad A_{\phi\,(L)} w^{LK}\\ \\w_{KL}(\Phi)A^{(L)}_\phi&\qquad\qquad w_{KL}\end{array} \right)
\end{equation}
\\
and the potential is
\begin{equation}\label{def of U}
    U(\Phi)=\frac{1}{F(r,\theta)}V(\Phi)
\end{equation}

It is seen that only the t-component of the new gauge fields is
non-zero. Using (\ref{1finitness of C anb B}) and (\ref{aa}), we see
that at the horizon $F_{\theta t}$ is zero for the both gauge
potentials.  From the definitions
 (\ref{definition of alpha}) and  (\ref{definition of beta}), we
 deduce that in  the case of the  double-horizon  we have,
 \begin{equation}\label{alpha beta}
 {\bf F}_{rt}\Bigg\vert_{r=r_H}\!\!\!\!\!\! =
 \left( \begin{array}{c} \alpha\\ \\ \beta^{(K)}
 \\\end{array}\right)
\end{equation}

In the three dimensional picture $C(r,\theta)$ becomes a component
of the new gauge filed. Because of the gauge freedom,  it is not
possible to determine this function by using the equations of
motion. This gauge fixing freedom of 3d picture is a consequence of
the reparametrization freedom of the original four dimensional
theory.

 Another  subtle point in the new 3d theory is that
coordinate $\theta$ goes from $0$ to $\pi$ and does not cover the
complete 3d space. In order to solve this problem, we use the axial
symmetry of the original solution and extend the new theory to the
whole of the space by demanding that
 \begin{equation}\label{parity symmetry1}
    X(\theta)=X(2\pi-\theta)
\end{equation}
where $X$ stands for any field  of   the 3d theory and $\theta$
covers the complete cycle  between $0$ to $2\pi$. In this form the
three dimensional theory is on the whole of $ R^3 $. Requirement of
the smoothness of the solution at the poles i.e $\theta=0,\pi$
implies ,
\begin{equation}\label{boundary condition}
    \frac{\partial X}{\partial\theta}\Bigg\vert_{\theta=0,\pi}\!\!\!\!\!=0
\end{equation}

Now we want  to impose    the double-horizon condition on the
equations of motion. Writing equations of motion at the horizon is
equivalent to removing all the  terms which have $S(r)$ factor.
Imposing the double-horizon condition is done by setting
 $\frac{dS(r)}{dr}$ factor to zero.
Variation of the metric in (\ref{three dimentinal action charged
case}) gives:
\begin{equation}
    R_{\mu\nu}-H^{ij}(\Phi)\partial_\mu\Phi_i
    \partial_\nu\Phi_j=W_{ab}(\Phi)\Big(2g^{\rho\sigma}F^a_{\mu\rho}F^b_{\nu\sigma}-g_{\mu\nu}F\,^a_{\rho\sigma}
    F^{b\,\,\rho\sigma}\Big)+ g_{\mu\nu}U(\Phi)
\end{equation}
 Setting $ S(r) $ and $\frac{dS(r)}{dr}$  to zero in this equation
 one obtains equations,
\begin{equation}\label{e.om for charged case from metric}
4N^2(\theta)+2\,N(\theta)\,M^{\prime\prime}(\theta)-N^\prime(\theta)\,M^\prime(\theta)+4\,N^2(\theta)\,M(\theta)\,U(\phi)=0
\end{equation}
\begin{equation}\label{e.om for charged case from metric2}
    4\,M(\theta)N(\theta)+\Big(M^\prime(\theta)\Big)^{2}=2\,M^2(\theta)H^{ij}(\Phi)\,\Phi^\prime_i\,\Phi^\prime_j+4\,N(\theta)\,V_e(\Phi)-2\,N(\theta)M^2(\theta)U(\Phi)
\end{equation}
where
\begin{equation}\label{def of Ve}
    V_e(\Phi)=B\big(W_{11}\,\alpha^2+\,W_{1K}\,\alpha\,\beta^{(K)}+\,W_{K1}\,\alpha\,\beta^{(K)}+\,W_{KL}\,\beta^{(K)}\beta^{(L)}\big)
\end{equation}
and derivatives are respect to $\theta$. Note that the $R_{\mu\nu}$
equations have resulted  only the  two above equations.
 The equations of motion from the variation of the scalars are;
\begin{equation}
    \frac{2}{\sqrt{-g}}\partial_\mu\Big(\sqrt{-g}H^{ij}(\Phi)\partial^\mu\Phi_j\Big)=\frac{\delta H^{kj}(\Phi)}{\delta
    \Phi_i}\partial_\mu\Phi_k\partial^\mu\Phi_j+\frac{\delta W_{ab}(\Phi)}{\delta
    \Phi_i}F^a_{\mu\nu}F^{b\,\mu\nu}+\frac{\delta U(\phi)}{\delta\Phi_i}
\end{equation}
which at the horizon of the  double-horizon black holes take the
form;
\begin{equation}\label{e.om for cgharegd for scalar 1}
    \frac{M(\theta)}{\sqrt{N(\theta)}}\Bigg(\frac{M(\theta)}{\sqrt{N(\theta)}}\,H^{ij}(\Phi)\Phi^\prime_j\Bigg)^\prime=\frac{M^2(\theta)}{2N(\theta)}\,\frac{\delta
    H^{jk}(\Phi)}{\delta\,\Phi_i}\,\Phi^\prime_j\,\Phi^\prime_k-\frac{\delta
    V_e(\Phi)}{\delta\,\Phi_i}+\frac{1}{2}M^2(\theta)\frac{\delta U(\phi)}{\delta\Phi_i}
\end{equation}

Finally the variation of the gauge fields gives;
\begin{equation}\label{variation of gauge CC}
    \partial_\mu\big(\sqrt{-g}\,W_{ab}(\Phi)\,F\,^{b\,\mu\nu}\big)=0
\end{equation}
The only non-trivial equation in (\ref{variation of gauge CC}) is
when $\nu=t$.  Using (\ref{gaugefilelds}) in this equation and
integrating with respect to $\theta$ in the interval $[0,\pi]$
gives;
\begin{equation}
   \frac{\partial}{\partial r}\Bigg(\int_0^\pi d\theta\,
   \sqrt{-g}\,g^{rr}g^{tt}\,W_{ab}(\Phi)\,F\,^b_{rt}\Bigg)+\Bigg[\sqrt{-g}\,g^{\theta\theta}g^{tt}\,W_{ab}(\Phi)\,F\,^b_{\theta t}\Bigg]_0^\pi=0
\end{equation}
  Using   (\ref{boundary condition}),  $F_{\theta t}=0$ at
   $\theta=0,\pi $, thus the second term  vanishes. The first term introduces a constant which for $a=1$ is
proportional to the angular momentum and for $a=K+1$ to the charge
$Q^{(K)}$ of the 4-dimensional black hole. At the horizon it takes
the form,
\begin{equation}\label{e.om for j}
   \sqrt{B} \int_0^\pi d\theta\,\frac{\sqrt{N(\theta)}}{M(\theta)}(W_{11}\alpha+W_{1K}\beta^{(K)})=-2J
\end{equation}

\begin{equation}\label{e.om for Q}
   \sqrt{B}  \int_0^\pi
     d\theta\,\frac{\sqrt{N(\theta)}}{M(\theta)}(W_{K1}\alpha+W_{KL}\beta^{(L)})=2Q^{(K)}
\end{equation}
where $J$ and $Q^{(K)}$ are the angular momentum and charges of the
black hole. The coefficients of proportionality at the right hand
sides are fixed   by use of  the  known  Kerr-Newman solution.

 Defining
\begin{equation}\label{lambda Phi}
      {\bf\Sigma}=\sqrt{B}\left(%
\begin{array}{c}
  \alpha \\ \\
  \beta^{(K)} \\
\end{array}%
\right),
\end{equation}
 the equations (\ref{e.om for charged case from metric}), (\ref{e.om for
charged case from metric2}), (\ref{e.om for cgharegd for scalar 1}),
(\ref{e.om for j}) and (\ref{e.om for Q}) take the following compact
forms,
 \begin{equation}\label{at1}
    4\,N^2(\theta)+2\,N(\theta)\,M^{\prime\prime}(\theta)-N^\prime(\theta)\,M^\prime(\theta)+4N^2(\theta)M(\theta)U(\Phi)=0
\end{equation}
\begin{equation}\label{at2}
    4\frac{N(\theta)}{M(\theta)}+\Big(\frac{M^\prime(\theta)}{M(\theta)}\Big)^2=2{\bf(\Phi^\prime)^T
    H\Phi^\prime} +4\,\frac{N(\theta)}{M^2(\theta)}\,{\bf\Sigma^T
    W\Sigma}-2N(\theta)U(\Phi)
\end{equation}
\begin{equation}\label{at3}
    \frac{\sqrt{N(\theta)}}{M(\theta)}\Bigg(\frac{M(\theta)}{\sqrt{N(\theta)}}\Big({\bf
    H\Phi^\prime}\Big)_i\Bigg)^\prime=\frac{1}{2}\,{\bf(\Phi^\prime)^T
    }\frac{\delta\bf{H}}{\delta\Phi_i}{\bf\Phi^\prime}-\frac{N(\theta)}{M(\theta)^2}{\bf(\Sigma)^T
    }\frac{\delta\bf{W}}{\delta\Phi_i}{\bf\Sigma}+\frac{N(\theta)}{2}\frac{\delta U(\Phi)}{\delta\Phi_i}
\end{equation}
\begin{equation}\label{at4}
    \int_0^\pi\,d\theta\,\frac{\sqrt{N(\theta)}}{M(\theta)}{\bf
    W\Sigma}=2{\bf Q}
\end{equation}
where
\begin{equation}\label{def of charg}
    {\bf Q}=\left(%
\begin{array}{c}
  -J \\ \\
  Q^{(K)} \\
\end{array}%
\right)
\end{equation}
Equations (\ref{at1})-(\ref{at4})  are a set of differential
equations which are decoupled from the bulk in the sense that  they
do not have any r-derivative term. They involve only functions of
angular variables and respective derivatives. Solving these
equations provides the  information only on  the horizon with no
reference to the bulk of the space and in particular asymptotic
values of the fields . This guarantees the decoupling of the
dynamics of the horizon from the bulk. If these equations have
unique solutions, this decoupling will prove the attractor mechanism
in its strong form. It implies that  the field configurations on the
horizon are determined uniquely by these equations and the behavior
at the infinity dose not enter in  specifying  their values. If the
equations admit more than one solution, the problem needs further
analysis and one must explore which class of asymptotic conditions
corresponds  to a particular  solution at the horizon
\cite{Goldstein:2005hq}.

It is notable  that solutions of these  equations determine
$M(\theta)$, $N(\theta)$, ${\bf\Phi}(\theta)$ and ${\bf\Sigma}$
which are required  for specifying the  near horizon geometry
(\ref{near horizon geometry of 4d-final form}). We find them without
using directly the equations in the near horizon. The boundary
values for solving these equations are given by (\ref{parity
symmetry1}) and (\ref{boundary condition}) .

 Our analysis clearly shows that this decoupling occurs for  the double-horizon black holes and for
distinct-horizon cases there is not such a decouplig. If we write
the field equations for the black holes with the distinct horizons ,
 a number of  terms with r-derivative factors will survive . The
 presence of
 these terms obstruct the decoupling of the equations and the
 behavior of the fields will depend on the boundary conditions at
 infinity. One expects that as the distance between the inner and outer
 horizon decreases and the black hole approaches the
 double-horizon case, the decoupling violating terms become smaller
  and correspondingly less important.

\subsection{Generalization to  f(R) Gravities}
Our method and argument   can be simply  generalized to include
$f(R)$ gravities. In these theories the term $\sqrt{-G}R$  in  the
action is replaced by the $ \sqrt{- G} f(R) $ resulting in the
action,
\begin{equation}\label{f(R) action}
     I=\frac{1}{\kappa^2}\int d^4 x
    \sqrt{- G} f(R)
\end{equation}
The other terms of the action remain unchanged and therefore we only
study the effect of this term on the above considerations. General
analysis of the  section two which is independent of any particular
action shows that the black hole solutions of this theory also have
the same form as (\ref{ansatz for the metric charged}).

The simplest way of  studying $f(R)$ gravities  is using  their
equivalence  to Einstein gravity coupled to a scalar field. If we
define $\sigma$ by
\begin{equation}\label{def of sigma}
    \sigma=\frac{\sqrt 3}{2}ln\Big\vert \kappa^2\,\frac{d\, f(R)}{dR}\Big \vert
\end{equation}
and make a conformal transformation,
\begin{equation}\label{f(R) conformal transformation}
    g_{\mu\nu}=\Big\vert \kappa^2\,\frac{d\, f(R)}{dR}\Big \vert\,G_{\mu\nu}=exp\Big(\frac{2\sigma}{\sqrt
    3}\Big)\,G_{\mu\nu}.
\end{equation}
The field equations derived from (\ref{f(R) action}) are equivalent
to those derived from the action
\begin{equation}\label{action hat}
    \hat S=\frac{1}{\kappa^2}\int d^4 x
    \sqrt{- g}\Big( R-2g^{\mu\nu}\partial_\mu\sigma
    \partial_\nu\sigma-V(\sigma)\Big)
\end{equation}
where
\begin{equation}\label{def of V}
    V=\lambda\,exp\Big(-\frac{4\sigma}{\sqrt
    3}\Big)\,\Big(R\frac{d\,f(R)}{dR}-f(R)\Big)
\end{equation}
and
\begin{equation}\label{def oflambda}
    \lambda=\Bigg\{\begin{array}{cc}
                     1 &if\quad \frac{d\,f(R)}{dR}>0 \\ \\
                     -1 &if\quad \frac{d\,f(R)}{dR}<0
                   \end{array}
\end{equation} It is seen that $V(\sigma)$ is the Legendre
transformation of $f(R)$.

Using (\ref{ansatz for the metric charged}) and (\ref{f(R) conformal
transformation}) we can write $g_{\mu\nu}$ as
\begin{equation}\label{gmunu}
    d\hat s^2=\hat A(r,\theta)\Bigg(-\frac{S(r)}{B(r,\theta)}\,dt^2+\frac{1}{S(r)}\,dr^2\Bigg)+\hat E(r,\theta)\,d\theta^2
    +\hat F(r,\theta)\Bigg(d\phi+C(r,\theta)\,dt\Bigg)^2
\end{equation}
where
\begin{eqnarray}
  \nonumber\hat A(r,\theta) &=& exp\Big(\frac{2\sigma}{\sqrt 3}\Big) A(r,\theta) \\
 \nonumber \hat E(r,\theta) &=& exp\Big(\frac{2\sigma}{\sqrt 3}\Big) E(r,\theta) \\
  \hat F(r,\theta) &=& exp\Big(\frac{2\sigma}{\sqrt 3}\Big) F(r,\theta)
\end{eqnarray} Hence the problem is reduced to the previous case
and we can repeat our method.

Reduction of this theory in $\phi$ direction results a theory with
the action
\begin{equation}\label{fr 3 action}
    \hat S=\frac{2\pi}{\kappa^2}\int d^3 x
    \sqrt{-\hat g}\Big(\hat R-2\hat g^{\mu\nu}\partial_\mu\Phi\partial_\nu\Phi-2\hat g^{\mu\nu}\partial_\mu\sigma\partial_\nu\sigma-w(\Phi)F_{\mu\nu}F^{\mu\nu}-U(\Phi,\sigma)\Big)
\end{equation}
where
\begin{eqnarray}
  w(\Phi) &=& \frac{1}{4}e^{-4\Phi} \\
  U(\Phi,\sigma) &=& e^{2\Phi}\,V(\sigma)
\end{eqnarray}
and
\begin{eqnarray}
  d\tilde s^2  &=&  M(r,\theta)\Bigg(-\frac{S(r)}{B(r,\theta)}\,dt^2+\frac{1}{S(r)}\,dr^2\Bigg)+  N(r,\theta)\,d\theta^2 \\
  e^{-2\Phi} &=&  \hat F(r,\theta)\\
  A_t &=&  C(r,\theta)
\end{eqnarray}
with
\begin{eqnarray}\label{MN in fr}
 \nonumber M(r,\theta) &=& \hat F(r,\theta)\,\hat A(r,\theta) \\
  N(r,\theta) &=& \hat F(r,\theta)\,\hat E(r,\theta)
\end{eqnarray}

 Field equations from variation of (\ref{fr 3 action}) with
respect to the metric, the scalars and the gauge filed are given by
\begin{equation}
    \hat R_{\mu\nu}-2\partial_\mu\Phi\partial_\nu\Phi-2\partial_\mu\sigma\partial_\nu\sigma=w(\Phi)(2F_{\mu\lambda}F_\nu\,^\lambda-\hat g_{\mu\nu}F_{\kappa\lambda}F^{\kappa\lambda})+\hat g_{\mu\nu}U(\Phi,\sigma)
\end{equation}
\begin{equation}
    \frac{1}{\sqrt{-\hat g}}\,\partial_\mu(\sqrt{-\hat g}\,\partial^\mu\Phi)=\frac{1}{4}\frac{\delta
w(\Phi)}{\delta\Phi}F_{\mu\nu}F^{\mu\nu}+\frac{1}{4}\frac{\delta
U(\Phi,\sigma)}{\delta\Phi}
\end{equation}
\begin{equation}
    \frac{1}{\sqrt{-\hat g}}\,\partial_\mu(\sqrt{-\hat g}\,\partial^\mu\sigma)=\frac{1}{4}\frac{\delta
U(\Phi,\sigma)}{\delta\sigma}
\end{equation}
\begin{equation}
    \partial_\mu(\sqrt{-\hat g}\,w(\Phi)\,F^{\mu\nu})=0
\end{equation}

 At the horizon
of the double-horizon black holes these equations take the forms,
\begin{equation}\label{fr1}
    4N^2(\theta)+2\,N(\theta)\,M^{\prime\prime}(\theta)-N^\prime(\theta)\,M^\prime(\theta)=-4\,N^2(\theta)\,M(\theta)\,U(\Phi,\sigma)
   \end{equation}
\begin{equation}\label{fr2}
   \begin{split} 4\,M(\theta)\,N(\theta)+\Big(M^\prime(\theta)\Big)^{2}=4M^2(\theta)\,(\Phi^\prime)^2+4M^2(\theta)\,(\sigma^\prime)^2 &+4\Sigma^2\,N(\theta)\,w(\Phi)\cr &-2\,N(\theta)\,M^2(\theta)\,U(\Phi,\sigma)\end{split}
\end{equation}
\begin{equation}\label{fr3}
\frac{M(\theta)}{\sqrt{N(\theta)}}\Bigg(\frac{M(\theta)}{\sqrt{N(\theta)}}\,\Phi^\prime\Bigg)^\prime=-\frac{1}{2}\,\Sigma^2\,\frac{\delta
w(\Phi)}{\delta\Phi}+\frac{1}{4}M^2(\theta)\,\frac{\delta
U(\Phi,\sigma)}{\delta\Phi}
\end{equation}
\begin{equation}\label{fr4}
\frac{M(\theta)}{\sqrt{N(\theta)}}\Bigg(\frac{M(\theta)}{\sqrt{N(\theta)}}\,\sigma^\prime\Bigg)^\prime=\frac{1}{4}M^2(\theta)\,\frac{\delta
U(\Phi,\sigma)}{\delta\sigma}
\end{equation}
\begin{equation}\label{fr5}
    \int_0^\pi d\theta\,\frac{\sqrt{N(\theta)}}{M(\theta)}\,\Sigma\,w(\Phi)\,=-2J
\end{equation}
The  derivatives are with respect to $\theta$ and we have defined
\begin{equation}
    \Sigma=\sqrt{B}\,\alpha
\end{equation}

Again these equations are a set of decoupled equations from the
bulk. Solving them  gives  $M(\theta)$, $N(\theta)$, $
\Phi(\theta)$, $ \sigma(\theta)$  and $\Sigma$ which determine the
near horizon geometry of the double-horizon black holes in $f(R)$
gravity.

 We note that a special case is when $f(R)= R + \Lambda $ where
 $\Lambda$ is a constant. In this case the black hole is
 asymptotically $AdS$. Hence  our analysis is valid even in
 the presence of the cosmological constant.

\section{Double-Horizon Limit and   Entropy Function \\ Method  }

Our Analysis in the previous sections clearly shows  the decoupling
of the dynamics at the horizon from the bulk. This decoupling occurs
at the level of the equations of motion, but an interesting question
is to see what happens  at the level of the action. We consider this
point in this section.

 Let us consider  charged rotating  black holes in the theory of
gravity which is described by the action (\ref{action }). The form
of  the metric, gauge field and scalar field are given by
(\ref{ansatz for the metric charged}), (\ref{ansatz for the scalar
scalar}) and (\ref{ansatz for the gauge filed 4d}). A proper way to
deal  with this theory is the reduction in $\phi$ direction and
studying the new three dimensional theory which is described by
(\ref{three dimentinal action charged case})-(\ref{definition of
W}). At the horizon of the double-horizon black holes, field
equations derived from the action of this new theory  take the forms
(\ref{at1})-(\ref{at4}).

We define
\begin{equation}\label{def L c}
    L= R-H_{ij}(\Phi)\partial_\mu\Phi^i\partial^\mu\Phi^j-W_{ab}(\Phi) F\,^a_{\mu\nu}
    F^{b\,\,\mu\nu}-U(\Phi)
\end{equation}
thus from (\ref{three dimentinal action charged case}) it is seen
that
\begin{equation}
    S=\frac{1}{8}\int d^3 x\,\sqrt{-g}  \,L
 \end{equation}
where we have chosen  $\kappa=16\pi$.

 By using the double-horizon condition, at the horizon we have

\begin{equation}\label{l c}
\begin{split}
    l(\theta)\equiv
    L\Bigg\vert_{r=r_H}\!\!\!\!\!\!\!=\frac{1}{2\,N^2(\theta)\,M^2(\theta)\,}\Bigg(& N(\theta)\,\big(M^\prime(\theta)\Big)^2+2\,M(\theta)\,N^\prime(\theta)\,M^\prime(\theta)-4\,N(\theta)\,M(\theta)M^{\prime\prime}(\theta)\cr &\,\,\,-4\,N^2(\theta)\,M(\theta)-2N(\theta)M^2(\theta){\bf(\Phi^\prime)^T
    H\Phi^\prime}\cr &+4N^2(\theta){\bf\Sigma^T W\Sigma}
    \Bigg)-U(\Phi)
\end{split}
\end{equation}
It is obvious that  $l(\theta)$ does not have any r-derivative term
and it is decoupled. Due to this  decoupled property of $l(\theta)$,
equations (\ref{at1})-(\ref{at4}) are given respectively as
\begin{equation}\label{f0 c}
    \frac{\delta f}{\delta M(\theta)}=0
\end{equation}
\begin{equation}\label{f01 c}
    \frac{\delta f}{\delta N(\theta)}=0
\end{equation}
\begin{equation}\label{f1 c}
    \frac{\delta f}{\delta{\Phi_i }}=0
\end{equation}
\begin{equation}\label{f2 c}
    \frac{\delta f}{\delta{\alpha_i}}={ Q_i}
\end{equation}
where
\begin{equation}
    f=\frac{1}{8}\int d\theta\,\sqrt {-g}\, l(\theta)
\end{equation}
Generating equations of motion  by extremizing a function which is
defined at the horizon, is similar to the  {\it {Entropy Function
Method }} which determines parameters of the near-horizon geometry
\cite{Sen:2005wa},\cite{Sen:2005iz}. The basic point here is that
in the our case this extremization is a direct consequence of the
double-horizon limit. It is not difficult to see that  $F$ which
is defined as
\begin{equation}\label{def entropy function}
    F=\alpha_i\,Q^i-f
\end{equation} is exactly
the entropy function if we started from the near horizon geometry.

Our analysis  shows that why we can not provide a method like the
entropy function method for the black holes with distinct horizons.
For these cases it is not possible to cancel r-derivative terms of
the action at the horizon and thus we do not have a decoupled action
which gives the equations of motion.

Using  equations (\ref{f0 c})-(\ref{f2 c}) we can simplify Wald's
entropy formula. We follow the method of \cite{Astefanesei:2006dd}.
The entropy of the  black hole is given by
\begin{equation}\label{5 wald}
    S_{BH}=-8\pi\,\sqrt{-h}\,\frac{\partial \mathcal{L}^{(2)} }{\partial
    R_{rtrt}^{(2)}}\,\sqrt{-h_{rr}h_{tt}}
\end{equation}
where $h_{\alpha\beta}$ with $\alpha,\beta\,=\, r,t$ is a two
dimensional metric  defined as
\begin{equation}\label{5 def of h}
    h_{\alpha\beta}=\frac{1}{2}\int_0^\pi\,d\theta\,sin\theta\,g_{\alpha\beta}
\end{equation}
and $\sqrt{-h}\,\mathcal{L}^{(2)} $ is the two dimensional
Lagrangian density, related to the three dimensional Lagrangian
density via the formula:
\begin{equation}\label{5 2d lagrangian }
    \sqrt{-h}\,\mathcal{L}^{(2)}=\int\,d\theta\sqrt{-g}\,\mathcal{L}
\end{equation}

It follows from (\ref{3d metric charged case}) and (\ref{5 def of
h}) that at the horizon, after imposing double-horizon limit we have
\begin{equation}\label{h and Rrtrt}
    \sqrt{-h_{rr}h_{tt}}=\sqrt{B}\,R_{rtrt}^{(2)}
\end{equation}
thus we can express (\ref{5 wald}) at the horizon as
\begin{equation}\label{5 wald b}
    S_{BH}=-8\pi\,\sqrt{B}\,\sqrt{-h}\,\frac{\partial \mathcal{L}^{(2)} }{\partial
    R_{rtrt}^{(2)}}\,R_{rtrt}^{(2)}
\end{equation}
Following the method of \cite{Astefanesei:2006dd}, one can use the
 equations (\ref{f0 c})-(\ref{f2 c})  to simplify (\ref{5 wald b})
and obtain
\begin{equation}\label{entropy}
    S_{BH}=2\pi\sqrt B\,\Big({\bf \alpha^T Q}-f\Big)
\end{equation}
where $f$ is evaluated at the extremized values. This is the same
result as the entropy function method. The extra factor $\sqrt B$ is
a result of this point that we did not use near-horizon geometry. We
can cancel it in the first term by the definition of $\Sigma$ and in
the second term by $\frac{1}{\sqrt B}$ factor of
 $\sqrt{-g}$, hence it  does not enter in the calculations.


%

\section{Conclusion}
The analysis provided in this paper has two aspects. First it gives
a deeper physical reason for  $AdS$ part of the near horizon
geometry of the double-horizon (extremal) black holes which has been
the beginning point of the entropy function method . Second it puts
our earlier result about the decoupling of the dynamics of the
horizon for double-horizon black holes on a firm ground. It also
opens venues  for further investigation of its properties.

Our analysis also clarify that why black holes with distinct
horizons do not enjoy a decoupling or attractor mechanism. The
nature of approach to the decoupling limit is also of interest. The
decoupling given as a set of equations on the compact sphere of the
horizon which can be solved consistently. If it has a unique
solution then the attractor mechanism works like the non rotating
case. Even if the solution is not unique we expect it to result in a
discrete set of solutions closely related to the minima of the
potential in the non-rotating case. The relation of different
solutions of  such discrete set to the large distance boundary
condition and quantum transition between these solution are not
clear yet.

 Solution to the set of the decoupled equations of the horizon
provides sufficient information for the physical properties of the
black hole in particular the entropy. Hence we may find a way to
understanding of the fact that all the information hidden in  a
black hole is distributed on the surface of the horizon.

The other interesting direction for further investigation is to
generalize this method to the higher dimensions where the topology
of the horizon is more complicated than a simple sphere.

These questions and other unclear properties of the double-horizon
limit is under investigation.

\section*{Acknowledgments}
We would like to thank M. Alishahiha ,  M. M. Sheikh Jabbari, A.
E. Mosaffa,  A. R. Tavanfar and A. Ghodsi for their useful
comments and discussions. We would also like to thank the Iranian
chapter of TWAS for partial support.


%
\end{document}